\documentstyle{article}

\def\gb{\beta}

\def\ge{\epsilon}

\def\gd{\delta}

\def\gm{\mu}

\def\gp{\pi}
\def\gP{\Pi}

\def\gs{\sigma}

\def\gl{\lambda}
\def\gL{\Lambda}

\def\go{\omega}

\def\delp{\partial_+}
\def\delm{\partial_-}

\def\delmu{\partial_\gm}

\def\delrlp{\stackrel {\leftrightarrow} {\partial_+}}
\def\delrlm{\stackrel {\leftrightarrow} {\partial_-}} 
\def\delrli{\stackrel {\leftrightarrow} {\partial_i}} 

\def\delrlmup{\stackrel {\leftrightarrow} {\partial^\gm}}
\def\part{\partial}
\def\parti{\partial_i}
\def\hlf{\frac{1}{2}}
\def\A0{A^{+}_0}

\def\xpl{x^{+}}
\def\xmin{x^{-}}
\def\ymin{y^{-}}
\def\xp{x_\perp}
\def\yp{y_\perp}
\def\ulix{\underline{x}}
\def\uliy{\underline{y}}

\def\ulin{\underline{n}}
\newcommand{\nc}{\newcommand}

\nc{\intl}{\int\limits_{-L}^{+L}\!dx^-}
\nc{\intv}{\int\limits_{V}^{}\!d^3\ulix}
\nc{\intvy}{\int\limits_{V}^{}\!d^3\uliy}
\nc{\intly}{\int\limits_{-L}^{+L}\!{{dy^-}\over\!2}}
\nc{\zmint}{\int\limits_{-L}^{+L}\!{{dx^-}\over{\!2L}}}
\nc{\intp}{\int\limits_{0}^{+\infty}\!{{dp^+}\over\!4\gp}}
\def\beq{\begin{equation}}
\def\eeq{\end{equation}}
\def\bea{\begin{eqnarray}}
\def\eea{\end{eqnarray}}
\begin{document}
\title{Higgs mechanism in a light front formulation}
\medskip
\author{ {\sl L$\!\!$'ubom\'{\i}r Martinovi\v c}
\footnote{E-mail: fyziluma@savba.sk}\\
Institute of Physics, Slovak Academy of Sciences\\  
D\'ubravsk\'a cesta 9, 845 11 Bratislava, Slovakia\\
\phantom{.}\\
and\\
\phantom{.}\\
{\sl Pierre Grang\'e}
\footnote{E-mail: Pierre.GRANGE@LPTA.univ-montp2.fr}\\
Laboratoire de Physique Th\'eorique et Astroparticules\\ 
Universit\'e Montpellier II, Pl. E. Bataillon\\ 
Montpellier, F-34095 France}
\medskip
\maketitle

\begin{abstract}
We give a simple derivation of the Higgs phenomenon  
in an abelian light front gauge theory. It is based on a finite-volume 
quantization with antiperiodic scalar fields and 
periodic gauge field. An infinite set of degenerate vacua 
in the form of coherent states of the scalar field, that minimize  
the light front energy, is constructed. The corresponding effective 
Hamiltonian describes a massive vector field whose third component is 
generated by the would-be Goldstone boson. This mechanism, 
understood here quantum mechanically in the form analogous to the spacelike 
quantization, is derived without gauge fixing as well as in the unitary 
and the light-cone gauge. 
\end{abstract}

\vspace{5mm}
Spontaneous symmetry breaking (SSB) of global as well as gauge symmetries 
has not been fully understood in the light-front (LF) field theory, which 
is the formulation (most often hamiltonian) of relativistic dynamics 
that uses the LF 
variables $x^\gm= (\xpl, \xmin,x^1,x^2), x^\pm=x^0 \pm x^3$ and is quantized 
on a surface of constant light-front time $\xpl$ \cite{Dir,LKS,Rohr}. 
Consequently, one deals with the LF field variables which satisfy field 
equations with a different structure than the equations of the conventional 
field theory which parametrizes the spacetime by  
$x^\mu = (t,x^1,x^2,x^3)$ and is quantized at $t=0$. 
The main reason for difficulties in obtaining a clear picture of SSB 
in the LF theory is the positivity of the spectrum of the LF momentum operator 
$P^+$.  Quite generally, the interacting-theory vacuum state coincides with 
the free Fock vacuum if Fourier modes carrying $p^+=0$ (LF zero modes - ZM) can 
be neglected. This simplifying feature is very useful in perturbative and 
bound-state calculations. However, it complicates the understanding of other 
non-perturbative properties because it seems to prohibit any vacuum structure 
in LF theories and hence also the well established SSB pattern. Alternative 
schemes of the physics of broken phase have been given in the LF literature 
\cite{lfssb,Yam,RT}. They are typically based on the operator scalar ZM 
which is present for periodic boundary conditions (BC) and which satisfies 
an equation of a constraint. The role of this  
variable in the phase transition of the $\gl \phi^4(1+1)$ model was 
analyzed by means of the Haag expansion in \cite{SGW}.   

The Higgs mechanism in the LF formalism was studied on the tree level in the 
continuum formulation \cite{Prem}. It was assumed that a scalar field contains 
a c-number piece which gave a justification for performing a usual shift 
in the Lagrangian leading to the generation of the mass term for the gauge 
field. A support for the above assumption comes from the fact that the solution 
of the zero-mode constraint of the real scalar field in the DLCQ analysis 
contains such a constant non-operator part \cite{Dave, LMSSB}. 

In the present work, we study the SSB of an abelian symmetry in the Higgs 
model. Our approach is based on the discrete light-cone quantization method 
(DLCQ) considered as a hamiltonian analytical framework with large but finite 
number of Fourier modes to approximate quantum field 
theory with its infinite number of degrees of freedom. A (regularized) unitary 
operator that shifts the scalar field by a constant will be used to transform  
the Fock space. The motivation for this step is a natural physical 
requirement to find ground states in the broken phase which would correspond 
to a lower LF energy than the usual Fock vacuum. This is suggested already 
by considering minima of the classical LF potential energy. A procedure, 
equivalent to transforming the states, is to work with a transformed 
Hamiltonian and calculate its matrix elements between the usual Fock states. 
In this way one naturally arrives at the effective type of the Hamiltonian 
that incorporates the usual pattern of the Higgs mechanism. 

The Lagrangian density of the abelian Higgs model that we wish to 
analyze has the form  
\beq
{\cal L}=-\frac{1}{4}F_{\mu\nu}F^{\mu\nu} + \hlf (D_\mu\phi)^\dagger D^\mu \phi + 
\hlf \gm^2 \phi^\dagger\phi - \frac{\gl}{4} (\phi^\dagger \phi)^2,  
\label{lagr}
\eeq
where $F_{\mu\nu}=\part_\mu A_\nu - \part_\nu A_\mu,~~D^\mu\phi=\part^\mu\phi 
+ieA^\mu\phi, \mu^2 > 0$. The Lagrangian is invariant under two groups of 
transformations: the global rotations of the complex scalar field $\phi(x) 
\rightarrow \exp{\big(-i\gb\big)}\phi(x)$ and the local gauge transformations 
\beq
\phi(x) \rightarrow \exp{\big(-i\go(x)\big)}\phi(x),~~A^\mu (x) \rightarrow  
A^\mu (x) + \part^\mu \go(x)/e.
\label{GT}
\eeq
In terms of the LF variables, the Lagrangian (\ref{lagr}) is 

\bea
&&\!\!\!\!\!\!\!\!\!\!\!\!{\cal L}_{lf} = 
\hlf \big(\delp A^+ - \delm A^-\big)^2 
+ \big(\delp A^i + \hlf \partial_i A^-\big)\big(2\delm A^i +\partial_i 
A^+\big) -  
\nonumber \\ 
&&\!\!\!\!\!\!\!\!\!\!\!\! - \hlf \big(\partial_1 A_2 - \partial_2 A_1)^2 +  
 \delp \phi^\dagger\delm\phi + \delm\phi^\dagger \delp\phi - 
\hlf \partial_i\phi^\dagger\partial_i\phi -  
\frac{ie}{2}\phi^\dagger\delrlp\phi A^+ - \nonumber \\ 
&&\!\!\!\!\!\!\!\!\!\!\!\! - \frac{ie}{2}\phi^\dagger\delrlm
\phi A^- - \frac{ie}{2}\phi^\dagger\delrli\phi A^i +  
\frac{e^2}{2}\big(A^+A^- - A^iA^i\big)\phi^\dagger\phi + \frac{\mu^2}{2}
\phi^\dagger\phi - \frac{\gl}{4}\big(\phi^\dagger\phi\big)^2.
\label{lflagr}
\eea 
Writing $\phi=\gs+i\gp$, the conserved current corresponding to 
the global symmetry is $J^\mu (x)=-i\phi^\dagger (x) \delrlmup \phi(x)=
2\gs (x)\delrlmup \gp (x)$. 
 
We will work in a finite volume $V=8L^3$ with space coordinates restricted to 
$-L \le \xmin,x^1,x^2 \le L$. Our notation is $x^\mu=(\xpl,\ulix), \ulix=
(\xmin,x^1,x^2),~ p.x=\hlf p^-x^+ + \hlf p^+x^- - p^1x^1-p^2x^2$. The gauge 
field will be chosen periodic in all three directions, while the scalar field 
will be antiperiodic: $A^\mu(\xpl,\xmin=-L, x,y)=A^\mu(\xpl,\xmin=L,x,y), 
\gs(\xpl,\xmin=-L,x,y)=-\gs(\xpl,\xmin=L, x,y)$, and similarly in the 
perpendicular directions $\xp \equiv(x^1,x^2)$ \cite{remark}.
The boundary conditions imply the discrete values 
of the three momentum labeled by a (half)integer and also lead  to the 
presence of global and proper zero modes of the gauge field \cite{Alex}. The 
proper ZM are constrained variables that can modify the LF Hamiltonian. For 
small coupling the corrections may be evaluated perturbatively \cite{AD}. 
We shall however neglect the gauge-field ZM in the present discussion 
because they are not crucial for the phenomenon under study. The 
fields below refer then to the sector of normal Fourier modes.

The LF Hamiltonian, obtained in the canonical way from the 
Lagrangian (\ref{lflagr}), reads
\bea
&&\!\!\!\!\!P^-=\intv \Big\{F_{12}^2 + \gP^2_{A^+}+2\gP_{A^+}\delm A^--\gP_{A^i}
\parti A^--  2e\gs\delrlm\gp A^-  
- 2e\gs\delrli \gp A^i -  \nonumber \\ 
&&\!\!\!\!\! - e^2 A^2\big(\gs^2+\gp^2\big)
+ (\parti\gs)^2 + (\parti \gp)^2 
- \mu^2\big(\gs^2+\gp^2\big) + 
\frac{\gl}{2}\big(\gs^2+\gp^2\big)^2 \Big\}. 
\label{Ham}
\eea
Here $A^2=A^+A^- - A^iA^i, i=1,2$ and the canonical momenta are 
\bea
&&\!\!\!\!\!\!\gP_{A^+}=\delp A^+-\delm A^-,~\gP_{A^i} = 
2\delm A^i + \parti A^+ 
\nonumber \\
&&\!\!\!\!\!\!\gP_{A^-}=0,~\gP_\gs=2\delm \gs-e\gp A^+,\gP_\gp=2\delm\gp+
e\gs A^+.
\label{moms}
\eea

At $\xpl=0$, we assume the usual LF commutation rules
\bea
&&\big[\gs(\xpl,\ulix),\gs(\xpl,\uliy)\big] = 
\frac{-i}{8}\ge(\xmin-\ymin)\gd^2(\xp-\yp), \nonumber \\
&&\big[\gp(\xpl,\ulix),\gp(\xpl,\uliy)\big] = 
\frac{-i}{8}\ge(\xmin-\ymin)\gd^2(\xp-\yp), \nonumber \\
&&\big[A^+(\xpl,\ulix),\gP_{A^+}(\xpl,\uliy)\big]=
\frac{i}{2}\gd^3(\ulix-\uliy) \nonumber\\
&&\big[A^i(\xpl,\ulix),\gP_{A^j}(\xpl,\uliy)\big]= 
\frac{i}{2}\gd^{ij}\gd^3(\ulix-\uliy).
\label{cr}
\eea
The mode expansions of the scalar fields are   
\bea
\gs(0,\ulix)=\frac{1}{\sqrt{V}}\sum\limits_{\ulin}^{}\frac{1}{\sqrt{p^+_n}}
\big[a(p_{\ulin})e^{-i p_{\ulin} .\ulix} + a^\dagger(p_{\ulin})e^{i p_{\ulin} .\ulix} \big], 
\nonumber \\
\gp(0,\ulix)=\frac{1}{\sqrt{V}}\sum\limits_{\ulin}^{}\frac{1}{\sqrt{p^+_n}}
\big[c(p_{\ulin})e^{-i p_{\ulin}. \ulix} + c^\dagger(p_{\ulin})e^{i p_{\ulin}.\ulix} \big],
\label{fexp}
\eea
where $p_{\ulin}=(p_n^+,p_{n_1},p_{n_2}), p_n^+=\frac{2\gp}
{L}n, n=1/2,3/2,\dots$, and similarly for the perpendicular components.     
The global rotations are implemented by the unitary operators $V(\gb)$ in terms of the charge $Q=\intv J^+(x)$:
\bea
\gs(x) \rightarrow V(\gb)\gs(x)V^\dagger(\gb) = \gs(x)\cos \gb-\gp(x)\sin \gb, \nonumber \\
\gp(x) \rightarrow V(\gb)\gp(x)V^\dagger(\gb) = \gs(x)\sin \gb+\gp(x)\cos \gb.
\label{ccr}
\eea
Here
\beq
V(\gb)=e^{i\gb Q}
\label{uniq}
\eeq
with 
\beq
V(\gb) = \exp\big\{\sum\limits_{\ulin}^{\gL}\Big(a^\dagger(p_{\ulin})
c(p_{\ulin}) - c^\dagger(p_{\ulin})a(p_{\ulin})\Big)\big\}. 
\label{glim}
\eeq
The Hamiltonian (\ref{Ham}) is invariant under $\xpl$-independent gauge 
transformations. They are implemented by the unitary operator
\beq
U[\go(\ulix)]=\exp \Big\{ i \intv\big[2\gP_{A^+}\delm - \gP_{A^i}\parti + 
eJ^+\big]\go(\ulix)\Big\} \label{loim}
\eeq
Indeed, we easily find 
\bea
&&U[\go(\ulix)]\phi(x)U^\dagger[\go(\ulix)]=\exp{\big(-i\go(\ulix)\big)}
\phi(x), \nonumber \\
&&U[\go(\ulix)]A^\mu(x)U^\dagger[\go(\ulix)]=
			A^\mu(x) + e^{-1}\partial^\mu\go(\ulix).
\label{gtimpl}
\eea
Consider now the unitary operators 
\bea
U_\gs(b)=\exp\Big\{-2ib\intv \gP_\gs(x)\Big\} \nonumber \\
U_\gp(b)=\exp\Big\{-2ib\intv \gP_\gp(x)\Big\}.
\label{shifto}
\eea
They shift the corresponding scalar field by a constant. To follow the usual 
treatment, we will perform only shifts in the $\gs$-direction:
\bea
U_\gs(b) \gs(x) U^{-1}_\gs(b) = \gs(x) - 2ib \intvy \big[\gP_\gs(y),\gs(x)\big] 
\nonumber \\
= \gs(x) - b \ge_\gL(L-\xmin) \ge_\gL(L-x^1) \ge_\gL(L-x^2).
\label{shift}
\eea
The subscript $\gL$ attached to the sign function $\ge(\ulix)$ indicates that their 
Fourier series is truncated at $\gL$:
\bea
\ge_\gL(\xmin) = \frac{4i}{L} \sum_{n=\hlf}^{\gL} \frac{1}{p_n^+} \Big(e^{-ip_n^+ \xmin} 
- e^{ip^+_n \xmin}\Big) 
\label{sign}
\eea
and analogously for the perpendicular components. The point is that one has to 
take a large but finite number of field modes in all three space directions in order 
to have a well-defined operator $U_\gs(b)$. In practice, for $\gL \approx 10^3$ the 
sign functions are equal to unity to a very good approximation everywhere on the finite 
interval $-L < \xmin,x^1,x^2 < L$ except for a very small neighborhood of the 
end-points. Therefore we 
will not write these sign functions explicitly henceforth. 

By means of the shift operator $U_\gs(b)$, we can define a set of states 
$\vert b \rangle = U_\gs(b)\vert 0 \rangle$ ($\vert 0 \rangle$ is the Fock 
vacuum). Minimizing the expectation value of the energy density $V^{-1}\langle 
b \vert P^- \vert b \rangle$, we easily find that the minimum of the LF energy, 
equal to $-\frac{\mu^4}{2\gl}$ is achieved for $b=\frac{\mu}{\sqrt{\gl}} \equiv 
v$. It is lower than the usual (vanishing) value of the LF energy in the 
``trivial'' vacuum $\vert 0 \rangle$. From Eq.(\ref{shift}) we also have the 
property that the vacuum expectation value of the $\gs$-field is non-zero 
which is the indication of broken symmetry: 
\beq
\langle v \vert \gs(x) \vert v \rangle = \langle 0 \vert U^{-1}_\gs(v) 
\gs(x)U_\gs(v) \vert 0 \rangle = v.
\label{ssb}
\eeq 
Here, the sign functions multiplying the value $v$ are understood as in 
Eq.(\ref{shift}). The accompanying vacuum degeneracy is easily obtained by 
rotating our ``trial'' vacuum (chosen in the $\gs$-direction):
\beq
V(\gb)\vert v \rangle = V(\gb)U_\gs(v)\vert 0 \rangle \equiv \vert v;\gb 
\rangle.
\label{fulv}
\eeq
Thus we have an infinite set of vacuum states corresponding to the above 
minimum of the LF energy and labeled by the angle $\gb$.

The next step in the Hamiltonian formalism is to construct the space of states. 
A natural possibility would be to apply a string of creation operators 
of all fields to the new vacuum, chosen to be $\vert v;0 \rangle$, and 
calculate the corresponding matrix elements of $P^-$. A simpler option is  
to build a usual set of Fock states from the Fock vacuum $\vert 0 \rangle$ 
and transform all of them by $U_\gs(v)$. This type of states is known as 
displaced number states in quantum optics \cite{qo}. In either case one can 
easily see that instead of the original Hamiltonian (\ref{Ham}) one actually 
works with the new ``effective'' LF Hamiltonian
\beq
\tilde{P}^- = U^{-1}_\gs(v) P^- U_\gs(v)
\label{Hef}
\eeq
in which the $\gs$-field is shifted by the value $v$. This of course leads to 
the structure known from the lagrangian formalism in the conventional field 
theory \cite{IZ}: the mass term of the gauge field of the form $e^2v^2A^2$ is 
generated, 
the pion field becomes massless and the $\gs$-field acquires mass equal to 
$\sqrt{2}\gm$. The change in the Hamiltonian density shows this explicitely: 
\bea
&&\gd P^- = -\frac{\gm^4}{2\gl} + 3\gm^2 \gs^2 + \gm^2 \gp^2 - e^2v^2A^2  
- 2e^2v\gs A^2 +   
\nonumber\\
&& + 2\sqrt{\gl}\gm \gs \Big(\gs^2 + \gp^2 \Big)
- 2ev\Big(\delm \gp A^- + \partial_i\gp A^i
\Big). 
\eea
The latter non-diagonal term and the kinetic term $(\parti \gp)^2$ can be 
removed by introducing the new field $B$:
\beq
B^{+}(x)=A^{+}+\frac{2}{ev}\delm\gp,~ 
B^i(x) = A^i(x) -\frac{1}{ev}\part_i \gp(x), 
\label{subst}
\eeq
while $B^- = A^-$. In this way, the $\gp$ field disappeared from the quadratic 
part of the Hamiltonian but it is still present in the interacting part. One 
may suspect that it is actually a redundant degree of freedom because 
the gauge freedom has not been removed.
  
In a full analogy with the space-like treatment, a clearer 
physical picture is obtained in the unitary gauge. Introducing the radial  
and angular field variables: 
\beq
\phi(x)=\rho(x)e^{i\Theta(x)/v},
\label{radial}
\eeq
the LF Hamiltonian will take the form
\bea
&&\!\!\!\!\!P^-_r=\intv \Big\{\gP^2_{A^+} + 2\gP_{A^+}\delm A^- - 
\gP_{A^i}\parti A^- 
+ F_{12}^2 +(\parti \rho)^2 + \nonumber \\
&&\!\!\!\!\! + \rho^2(\parti \Theta/v)^2  - 2e\rho^2\delm 
A^-\Theta/v  - 2e\rho^2A^i \parti \Theta /v -e^2\rho^2 A^2 -\mu^2\rho^2 + 
\frac{\gl}{2}\rho^4 \Big\}. 
\label{hamr}
\eea
To fix the gauge at the classical Lagrangian level, one observes that the gauge 
transformations simply shift the angular field variable $\Theta(x)$ by the 
gauge function $\go(x)$. Choosing $\go(x)=-\Theta(x)/v$, one has 
\beq
\! \phi(x) \rightarrow \rho(x),~A^\mu (x) \rightarrow B^\mu (x) = A^\mu (x) - 
\frac{1}{ev}\part^\mu \Theta (x)
\label{newa}
\eeq
with the corresponding Lagrangian
\beq
{\cal L}_u = -\frac{1}{4}G_{\mu \nu}G^{\mu \nu} + \hlf |\delmu \rho 
-ieB_\mu \rho|^2 + \hlf\mu^2 \rho^2 
-\frac{\gl}{4}\rho^4.
\label{ulag}
\eeq 
Taking this gauge fixing over to the quantum theory, defined by the 
commutation relation at $\xpl=0$
\beq
\Big[\rho(\xpl,\ulix),\rho(\xpl,\uliy)\Big] = -\frac{i}{8}
\ge(\xmin-\ymin)\gd^2(\xp-\yp),
\label{crr}
\eeq
we find the quantum LF Hamiltonian $P^-_u$ in the unitary gauge. It coincides 
with the Hamiltonian (\ref{hamr}) except for the missing $\Theta$-terms and 
the $B^\mu$ replacing the $A^\mu$ field. The equal-LF time algebra (\ref{crr}) 
enables us to introduce the shift operator ($\gP_\rho = 2\delm \rho)$
\beq
U_\rho(v)=\exp\Big\{-2i v\intv \gP_\rho(x) \Big\}
\label{shifr}
\eeq
which defines the ``effective'' LF Hamiltonian 
$\tilde{P}^-_u = U^{-1}_\rho (v) P^-_u U_\rho(v)$ 
corresponding to the unitary gauge:
\bea
&&\!\!\!\!\!\!\tilde{P}^-_u = \intv \Big\{\gP_{B^+}^2 + 2\gP_{B^+}\delm B^- 
- \gP_{B^i}\partial_i B^- + G_{12}^2 + ~~~~  
\nonumber \\ 
&&\!\!\!\!\!\!+ (\partial_i\rho)^2   
-e^2(\rho + v)^2B^2 - \mu^2(\rho + v)^2 + \frac{\gl}{2}(\rho + v)^4\Big\}. 
\label{Hunit}
\eea
From its form it is easy to find that it describes one massive scalar field 
$\rho$ and a vector field with the mass $e^2v^2$. The massive vector field 
emerged as a combination of the the massless gauge field $A^\mu$ and the 
scalar $\Theta$ field. 

Another possibility is to analyze the symmetry breaking in the light-cone 
gauge. This means that we set $A^+ = 0$ in the normal-mode sector. The 
starting Hamiltonian and conjugate momenta are then the expressions 
(\ref{Ham}),(\ref{moms}) without the terms containing $A^+$. One proceeds 
as in the case without the gauge fixing, namely defines the 
shift operator $U_{\gs}(v)$ and constructs the infinite set of degenerate 
(approximative) vacuum states by applying the unitary operator $V(\gb)$ 
(Eq.(\ref{glim})) to the coherent-state vacuum $\vert v \rangle$. The 
corresponding effective LF Hamiltonian is obtained by the transformation 
(\ref{Hef}). One observes an important difference as compared 
with the unitary-gauge treatment. It is related to the fact that the choice  
$A^+ = 0$ eliminates the $A^+ A^-$ part of the vector field mass term 
generated by shifting the $\gs$ field in the $-e^2 A^2 (\gs^2 + \gp^2)$ term 
in the Hamiltonian (\ref{Ham}). Thus the massive vector field seems to have 
only two components and this is not correct. The resolution of this difficulty 
comes from the observation \cite{Prem} that in the light-cone gauge the 
Gauss' law becomes a constrained equation for the $A^-$ component of the 
gauge field: 
\beq
\delm^2 A^-(x) + \delm \parti A^i(x) = e\gs(x) \delrli \gp(x). 
\label{amcon}
\eeq
The shift of the $\gs$ field by means of the operator $U_\gs(v)$ generates 
an additional term of the form $ev\delm \gp(x)$ on the righ-hand side of this 
equation. Upon inserting the shifted constraint to the Hamiltonian, the latter 
piece leads to the new term $e^2v^2 \gp^2$ ($i=1,2$): 
\beq
\tilde{P}^-_{lc}= \intv \Big[F_{12}^2 + (\parti A^i)^2 + (\parti \gp)^2 
+ e^2v^2 \big(\gp^2 + A_i^2\big) + \dots \Big].
\label{frep}
\eeq
To see that this Hamiltonian corresponds to a free massive vector meson field, 
it is useful to consider the gauge-invariant form of the scalar electrodynamics with a massive vector field \cite{Sop}. It differs from the Lagrangian 
of the massless scalar QED by the term $\hlf(mA^\gm-\partial^\gm B)^2$ which 
makes the vector field massive. $B$ is a scalar field and $m$ a mass parameter. The usual formulation with the condition $\partial_\gm A^\gm = 0$ corresponds 
to the gauge $B=0$. In the $A^+=0$ gauge we obtain  
\beq
P^- = \intv  \Big[F_{12}^2 + \big(\parti A^i)^2 + (\parti B)^2 + m^2 A_i^2 
+m^2B^2\Big],   
\label{lfmvbh}
\eeq 
plus the interaction terms. Comparing the two Hamiltonians, one can see that 
also in the light-cone gauge picture of the LF Higgs mechanism the gauge field 
became massive possessing three components $(\gp,A^1,A^2)$ with the mass 
$m=ev$. The mass term of the $\gs$ field is generated as in the previous case.  
 
In summary, we gave three versions of the Higgs phenomenon in the light front 
abelian Higgs model for different gauge choices. Our light front formulation 
was based on the finite-volume quantization with antiperiodic boundary 
conditions for the scalar fields. Minimization of the LF energy led 
to the semiquantum description of the degenerate vacuum states. In this way, 
the concept of the trivial LF vacuum containing no quanta was generalized to 
a more complex vacuum state with the non-trivial structure. The overall picture of the spontaneous breaking of the (abelian) gauge symmetry was thus found to 
be quite analogous to the conventional theory quantized on the space-like 
hypersurface, namely one scalar field and the gauge field become massive 
(the tree-level masses $e^2v^2$ and $\sqrt{2}\mu$, respectively) and there 
is no massless Goldstone boson in the particle spectrum.

This work was supported by the grant No. APVT 51-005704 of the Slovak Research  
and Development Agency, by the French NATO fellowship and by 
IN2P3-CNRS. Hospitality of the LPTA Laboratory at the Montpellier University 
is also gratefully acknowledged by L.M..
 

\begin{thebibliography}{99}
\bibitem{remark}
The alternative 
periodic BC for scalar fields imply presence of the constrained zero mode 
which in some cases lead to a non-vanishing vacuum expectation value (VEV) 
of a field\cite{lfssb,Yam}. We believe however that the overall physical 
picture of the symmetry breakdown is considerably clearer with the choice of 
antiperiodic BC.
\end{thebibliography}
\end{document}